\documentclass[twocolumn,aps,superscriptaddress,showpacs,nofootinbib,floatfix]{revtex4-1}
\usepackage{tabstackengine}
\usepackage{tikz}
\usepackage{tikz-3dplot}
\tdplotsetmaincoords{80}{45}
\tdplotsetrotatedcoords{-90}{180}{-90}

\tikzset{surface/.style={draw=blue!70!black, fill=blue!40!white, fill opacity=.6}}

\usepackage{epsfig,bm}
\usepackage{graphicx,amssymb}
\usepackage{amsmath}
\usepackage{mathtools}
\usepackage{physics}
\usepackage{tensor}
\usepackage{diagbox}
\usepackage{epstopdf}
\usepackage{float}
\usepackage{tabularx}
\usepackage{subfigure}
\usepackage{soul}
\usepackage{tikz-feynman}
\tikzfeynmanset{compat=1.0.0}

\usepackage[eps]{pstricks}
\usepackage[normalem]{ulem}  % 

\renewcommand\sout{\bgroup \color{red} \ULdepth=-.5ex \ULset}
\usepackage{slashed}
\newcommand{\comment}[1]{}
\begin{document}

%\title{General angular distribution of $\phi\to h+\bar{h},\;\phi\to l+\bar{l}$ channel and its dependence on polarization of $\phi$ meson}
\title{Identifying the transverse and longitudinal modes of the $K^*$ and $K_{1}$ mesons through their angular dependent decay modes}

\author{In Woo Park}%
\email{darkzero37@naver.com}
\affiliation{Department of Physics and Institute of Physics and Applied Physics, Yonsei University, Seoul 03722, Korea}

\author{Hiroyuki Sako}
\email{hiroyuki.sako@j-parc.jp}
\affiliation{Advanced Science Research Center, Japan Atomic Energy Agency, Tokai, Naka, Ibaraki 319-1195, Japan}

\author{Kazuya Aoki}
\email{kazuya.aoki@kek.jp}
\affiliation{KEK, High Energy Accelerator Research Organization, Tsukuba, Ibaraki 305-0801, Japan}

\author{Philipp Gubler}
\email{philipp.gubler1@gmail.com}
\affiliation{Advanced Science Research Center, Japan Atomic Energy Agency, Tokai, Naka, Ibaraki 319-1195, Japan}

\author{Su Houng Lee}%
\email{suhoung@yonsei.ac.kr}
\affiliation{Department of Physics and Institute of Physics and Applied Physics, Yonsei University, Seoul 03722, Korea}

%%%%%%%%%%%%%%%%%%%% Abstract %%%%%%%%%%%%%%%%%%%%%

%\begin{abstract}
%We use phenomenological interaction lagrangian to calculate the general angular distribution of the $\phi$ meson decaying through hadronic channel and leptonic channel. The system is found out to be anisotropic. However, azimuthal dependence can be averaged over and cancelled out, so that we can discriminate over transverse and longitudinal polarization. At specific angle, we can solely pick up transverse polarization or longitudinal polarization. Also, we can observe how decay amplitude for each polarization is distributed over momentum of daughter particle measured in the lab frame.
%\end{abstract}
\begin{abstract}
Observing the mass shifts of chiral partners will provide invaluable insight into the role of chiral symmetry breaking in the generation of hadron masses. Because both the $K^*$ and $K_1$ mesons have vacuum widths smaller than 100 MeV, they are ideal candidates for realizing mass shift measurements. On the other hand, the different momentum dependence of the longitudinal and transverse modes smear the peak positions. In this work, we analyze the angular dependence of the two-body decays of both the $K^*$ and $K_1$. It is found that the longitudinal and transverse modes of the $K^*$ can be isolated by observing the pseudoscalar decay in either the forward or perpendicular directions, respectively. For the $K_1$ decaying into a vector meson and a pseudoscalar meson, one can accomplish the same goal by further observing the polarization of the vector meson through its angular dependence on the two pseudoscalar meson decay. 
\end{abstract}

\maketitle

\section{Introduction}\label{sec:introduction}

Understanding the generation of hadron masses stands as one of the fundamental puzzles in Quantum Chromodynamics (QCD).  It is widely believed that spontaneous chiral symmetry breaking \cite{Nambu:1961tp,Nambu:1961fr} partly contributes to the generation of hadronic  masses 
\cite{Hatsuda:1985eb,Brown:1991kk,Hatsuda:1991ez,Leupold:2009kz}.  
Experiments conducted worldwide have aimed to observe the mass shift of hadrons at finite temperatures or densities \cite{Hayano:2008vn,JPARC:2023quf,Metag:2017yuh,Ohnishi:2019cif,Salabura:2020tou}.
This is because chiral symmetry is expected to be partially restored in the initial stages of relativistic 
heavy ion collisions and in nuclear matter probed by nuclear target experiments, respectively.

In particular, the J-PARC E16 experiment \cite{JPARC:2023quf, Aoki:2023qgl} will pursue the observation of the mass shift of the $\phi$ meson through $e^+e^-$ pairs emanating from pA collisions. This measurement will be complemented by the J-PARC E88 experiment \cite{Sako}, which aims to measure the $\phi$ meson through its $K^+K^-$ decay. The $\phi$ is expected to be a particularly sensitive probe, 
as its vacuum width is small, meaning that any width increase in the medium will not be significant enough to disrupt experimental reconstruction of the peak position\,\cite{Gubler:2024day}.

On the other hand, to isolate the effect of chiral symmetry restoration in a medium, the transformation of chiral partners towards degeneracy 
would be a critical experimental signal. This inevitably leads us to study the $K^*,K_1$ system as they appear to be the only realistically observable chiral partners, of which both have small vacuum widths \cite{Lee:2019tvt,Song:2018plu}.

The existence of the spin degrees of freedom, however, makes the situation more complicated, as both vector and axial vector mesons 
will have different responses depending on their spin orientation with respect to their motion relative to the medium. This effect is dominated by non-chiral symmetry-breaking effects \cite{Lee:2023ofg}, but it will cause the longitudinal and transverse modes to diverge for larger momenta, obscuring the peak position\,\cite{Lee:1997zta,Kim:2019ybi}.

In a recent publication \cite{Park:2022ayr}, we have shown that the longitudinal and transverse modes of the $\phi$ meson can be discriminated by analyzing the angular dependence of its two-body decay. In particular, the $e^+e^-$ and $K^+K^-$ decays can be used as complementary measurements.

In this work, we analyze the angular dependence of the two-body decays of both the $K^*$ and $K_1$. As we will show, the longitudinal and transverse modes of the $K^*$, or any other vector meson such as the $\rho$ (which decays from $K_1$) can be isolated by observing the pseudoscalar decay in either the forward or perpendicular directions, respectively. For the $K_1$ decaying into a vector meson and a pseudoscalar meson, one can accomplish the same goal by further observing the polarization of the vector meson as discussed before. 

The paper is organized as follows. In Sec.\ref{sec:decayrate}, we introduce the relevant effective interaction Lagrangians and estimate the coupling constants for each decay. Then we study a spin-1 particle state with a superposition of three different helicities and discuss how the general angular distribution is connected to the spin density matrix. We furthermore point out that the same result can be obtained using the helicity formalism. We then summarize our discussion in Section.\ref{sec:sumandconc}. More details regarding the calculations are provided in the appendices.

\section{(Axial)Vector meson decay rate}\label{sec:decayrate}

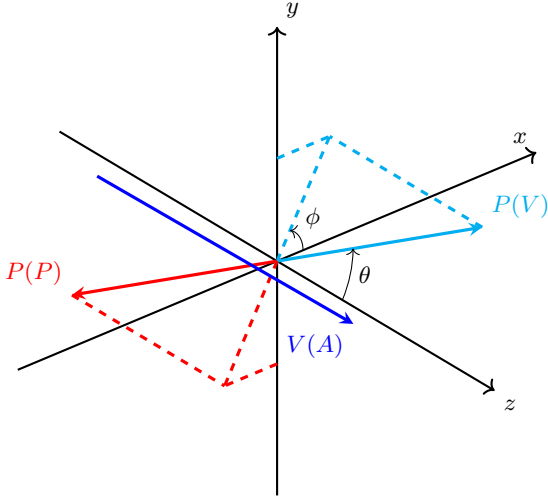
\begin{figure}[H]
\tdplotsetmaincoords{60}{110}
\hspace*{0.3cm}
\begin{tikzpicture}
[scale=4.5,tdplot_main_coords,rotate around z=0,rotate around x=-90,rotate around z=180,rotate around y=-30]
 % variables
  \def\rvec{1}
  \def\thetavec{30}
  \def\phivec{60}
 
  % axes
  \coordinate (O) at (0,0,0);
  \draw[thick,->] (-1.0,0,0) -- (1.0,0,0) node[above left]{$x$};
  \draw[thick,->] (0,-0.8,0) -- (0,0.8,0) node[above right]{$y$};
  \draw[thick,->] (0,0,-1.0) -- (0,0,1.0) node[below right]{$z$};

\tdplotsetcoord{P}{0.8\rvec}{\thetavec}{\phivec}
  \draw[-stealth,very thick,cyan] (O)  -- (P) node[above right] {$P(V)$};
  \draw[dashed,very thick,cyan]   (O)  -- (Pxy);
  \draw[dashed,very thick,cyan]   (P)  -- (Pxy);
  \draw[dashed,very thick,cyan]   (Py) -- (Pxy);
  
\tdplotsetcoord{Q}{0.8\rvec}{-\thetavec+180}{\phivec+180}
  \draw[-stealth,very thick,red] (O)  -- (Q) node[above left] {$P(P)$};
  \draw[dashed,very thick,red]   (O)  -- (Qxy);
  \draw[dashed,very thick,red]   (Q)  -- (Qxy);
  \draw[dashed,very thick,red]   (Qy) -- (Qxy);
%\tdplotdrawarc[coordinate system, draw styles]{center}{r}{angle start}{angle end}{label options}{label}

  \tdplotdrawarc[->]{(0,0,0)}{0.1}{0}{\phivec}{above=2pt,right=-1pt,anchor=south west}{$\phi$}
  
  \tdplotdrawarc[->,rotate around z=\phivec-90,rotate around y=-90]{(0,0,0)}{0.3}{0}{\thetavec}
    {anchor=west}{$\theta$}

\tdplotsetcoord{R}{0.6\rvec}{180-26.5}{\phivec+90}

\tdplotsetcoord{S}{0.6\rvec}{20.5}{\phivec+90}
\draw[-stealth,very thick,blue] (R)  -- (S) node[below left] {$V(A)$};
\end{tikzpicture}
\caption{Spin-1 particle decay in its rest frame. Cyan and red arrows each denote one decay particle after decay. The blue arrow stands for the traveling direction of the (axial) vector-meson in the Lab frame.}
\label{fig:kinematics}

\end{figure}

In this section, we will introduce the basic kinematics of the two-body decay channels, along with phenomenological Lagrangians describing the interactions between the relevant particles, and estimate the corresponding hadronic coupling constants.  We closely analyze the decay channels $\rho(770)\to\pi\pi$ and $K^{*}(892)\to K\pi$, both of which will be denoted as $V\to PP$. For the $A\to VP$ decays, we study $K_{1}(1270)\to\rho(770)K$ and $K_{1}(1270)\to K^{*}(892)\pi$. Here $P$ denotes a pseudoscalar meson while $V$ and $A$ denote a vector meson and an axial vector meson, respectively.
In the decay, we will denote $\theta$ and $\phi$ as polar and azimuthal angles of one of the decay products, measured in the center of mass (c.m.) frame (see Fig.\,\ref{fig:kinematics}). The z-axis is defined to align with the momentum direction of the initial particle in the Lab frame.

We assume that the initial (axial)vector meson is a superposition of the different helicity states $\ket{\lambda}$ ($\lambda = \pm 1$: transverse polarization, $\lambda = 0$: longitudinal polarization) with respective amplitudes $a_{\lambda}$. We can hence express the general (axial)vector meson state as
\begin{align}
   \ket{V/A} = \displaystyle\sum_{\lambda=\pm1,0}a_{\lambda}\ket{\lambda}. 
   \label{eq:vec_meson_config}
\end{align}
The spin density matrix $\rho_{\lambda\lambda^{\prime}}$ is defined using the coefficients $a_{\lambda}$ and reads
\begin{align}   \rho_{\lambda\lambda^{\prime}}=a_{\lambda}a_{\lambda^{\prime}}^{\star}.
\label{eq:spin_density_def}
\end{align}
The trace of the spin density matrix is normalized to 1: $\rho_{11}+\rho_{00}+\rho_{-1-1}=1$. 
For a transversely polarized (axial)vector meson, the meson spin $z$ component will be $J_{z}=\pm1$, thus $\rho_{00}=0$. In contrast, if the meson is 
longitudinally polarized, $J_{z}=0$ and $\rho_{00}=1$. The density matrix of an unpolarized meson has diagonal entries of 1/3, specifically $\rho_{11}=\rho_{00}=\rho_{-1-1}=\frac{1}{3}$.

%\begin{figure}[H]\centering\begin{tikzpicture}\begin{feynman}[large]\vertex (a) ;\vertex [right=of a] (b);\vertex [above right=of b] (f1){\textcolor{cyan}{\(P(V)\)}};\vertex [below right=of b] (f2){\textcolor{red}{\(P(P)\)}};\diagram* {(a) -- [blue,edge label'=\textcolor{blue}{\(V(A)\)},fermion,momentum={[arrow style=blue]\(q\)}] (b) , (b) -- [cyan,fermion,momentum={[arrow style=cyan]\(p_{1}\)}] (f1), (b) -- [red,fermion,momentum={[arrow style=red]\(q-p_{1}\)}] (f2)};\end{feynman}\end{tikzpicture}\label{fig:phitokaon}\caption{Tree-level diagram of a (axial) vector-meson decay}\label{fig:feyndiagram}\end{figure}%

\subsection{$\rho\to\pi\pi$ and $K^{*}\to K\pi$ decay}\label{sec:vectordecay}

The phenomenological interaction Lagrangians of the vector meson with two pseudoscalar mesons used in this work, are adapted from Ref.\cite{Sung:2021myr} and given as 
\begin{align}
&\mathcal{L}=g_{\rho\pi\pi}\bigg(\pi^{+}\overset{\leftrightarrow}{\partial}_\mu\pi^{-}\rho_{0}^{\mu}+\pi^{+}\overset{\leftrightarrow}{\partial}_\mu\pi^{0}\rho_{-}^{\mu}+\pi^{-}\overset{\leftrightarrow}{\partial}_\mu\pi^{0}\rho_{+}^{\mu}\bigg),\\
&\mathcal{L}=\sqrt{2}g_{K^{*}K\pi}\bigg(\bar{K}\vec{\tau}\vdot\partial_{\mu}\vec{\pi}-\partial_{\mu}\bar{K}\vec{\tau}\vdot\vec{\pi}\bigg)K^{*\mu}.
\label{eqn:kaon_lagrangian}
\end{align}
$K^{*},\,K$ are isomultiplets,  their matrix representation being listed in  Appendix\,\ref{appendix:effectivelagrangian}. All the masses of isomultiplets are isospin averaged using the PDG data \cite{Workman:2022ynf}, giving $m_{K}=495.644$ MeV, $m_{K^{*}}=893.61$ 
MeV, $m_{\rho}=775.16$ MeV and $m_{\pi}=138.037$ MeV. Similarly, in order to evaluate the coupling constants $g_{\rho\pi\pi}$ and $g_{K^{*}K\pi}$, we use the partial decay width from the PDG \cite{Workman:2022ynf}. The initial spin average involves a total of 3 degrees of freedom. 
For $\rho\to\pi\pi$, depending on the isospin, the decay modes are $\rho^{+}\to\pi^{+}\pi^{0},\,\rho^{-}\to\pi^{-}\pi^{0}$ and $\rho^{0}\to\pi^{+}\pi^{-}$. 
For the $K^{*}\to K\pi$ decay, they are $K^{+*}\to K^{+}\pi^{0},\,K^{0}\pi^{+}$ and $K^{0*}\to K^{0}\pi^{0},\,K^{+}\pi^{-}$. 
Therefore, after summing over the initial isospin components, the average is obtained by dividing by a factor of 3 for $\rho\to\pi\pi$ and 4 for $K^{*}\to K\pi$. The respective widths are then obtained as
\begin{align}
\label{eq:decaywidth}
\Gamma_{\rho\pi\pi}&=\frac{g_{\rho\pi\pi}^{2}}{8\pi}\frac{\abs{\bm{p_{1}}}}{m_{\rho}^{2}}\frac{4}{3}\abs{\bm{p_{1}}}^{2}=149\,\text{MeV},\nonumber\\
&\\ \nonumber
\Gamma_{K^{*}K\pi}&=\frac{g_{K^{*}K\pi}^{2}}{8\pi}\frac{\abs{\bm{p_{1}}}}{m_{K^{*}}^{2}}4\abs{\bm{p_{1}}}^{2}=51.4\,\text{MeV}, 
\end{align}
where 
\begin{align}\abs{\bm{p_{1}}}&=\frac{1}{2m_{V}}\sqrt{\big(m_{V}^{2}-(m_{1}-m_{2})^{2})\big(m_{V}^{2}-(m_{1}+m_{2})^{2}\big)} \nonumber
\end{align}
is the momentum of the two produced particles in the c.m. frame, while $m_{V}$ stands for the mass of the initial particle. For the $V\to PP$ decay, $m_{1}$ is taken to be one of the outgoing $\pi$ mesons. For the $A\to VP$ decay, $m_{1}$ is the mass of the produced vector-meson. 
From the partial decay width of the initial vector-meson, we can obtain the coupling strength of each decay channel. The resultant coupling constants of the respective interaction Lagrangians are listed in Table\,\ref{table:moment_coupl}.

Assuming that the initial vector-meson is in the general configuration of Eq.\,(\ref{eq:vec_meson_config}), we can obtain the general angular distribution as \cite{Schilling:1969um} 
\begin{align}
\frac{1}{\Gamma}\dv{\Gamma}{\Omega}&=\frac{3}{8\pi}\bigg(2\rho_{00}\cos^{2}\theta+(1-\rho_{00})\sin^{2}\theta\nonumber\\
&-2\text{Re}[\rho_{1-1}]\sin^{2}\theta\cos2\phi+2\text{Im}[\rho_{1-1}]\sin^{2}\theta\sin2\phi\nonumber\\
&-\sqrt{2}\text{Re}[\rho_{10}-\rho_{-10}]\sin2\theta\cos\phi\nonumber\\
&+\sqrt{2}\text{Im}[\rho_{10}+\rho_{-10}]\sin2\theta\sin\phi\bigg),
\label{eq:vpp_angle_dist}
\end{align}
where $\theta$ and $\phi$ are as before the polar and azimuthal angles of the outgoing daughter particle. The details of this calculation are given in the Appendix \ref{appendix:polarizationtensor}. Integrating $\frac{1}{\Gamma}\dv{\Gamma}{\Omega}$ over $\phi$, we acquire the polar distribution $W(\theta)$ as
\begin{align}
W(\theta)&=\frac{3}{4}\bigg((1-\rho_{00})+(3\rho_{00}-1)\cos^{2}\theta\bigg).\label{eq:Wthetaphi}
\end{align}
If we substitute $\rho_{00}=0$, $W(\theta)$ becomes the decay distribution of a transversely polarized vector meson, while for $\rho_{00}=1$, we get its longitudinal counterpart. These results agree with the result derived using polarization tensor \cite{Park:2022ayr}.

%\begin{widetext}

\begin{table}[t]
\caption{Coupling constant for each decay channel and the respective momentum of the daughter particle in the c.m. frame.}
\centering
 \begin{tabular}{ c | c | c | c | c }
    \hline
    Decay & $\rho\to\pi\pi$  & $K^{*}\to K\pi$ &  $K_{1}\to\rho K$ & $K_{1}\to K^{*}\pi$
    \\
    \hline
    $\abs{\bm{p_{1}}}$(MeV) & $362$   & $289$ &   $27$ & $299$    \\
   \hline
  $g_{ABC}$ &  $5.96$  &  $3.27$  &  $3.26$  &  $0.71$  \\
  \hline
\end{tabular}
\label{table:moment_coupl}
\end{table}

%\end{widetext}

\begin{figure}
\centering
\subfigure [$\dv{\Gamma}{\cos\theta}$ of $\rho\to\pi\pi$ in the c.m. frame]{
\includegraphics[width=3.3in,height=1.7in]{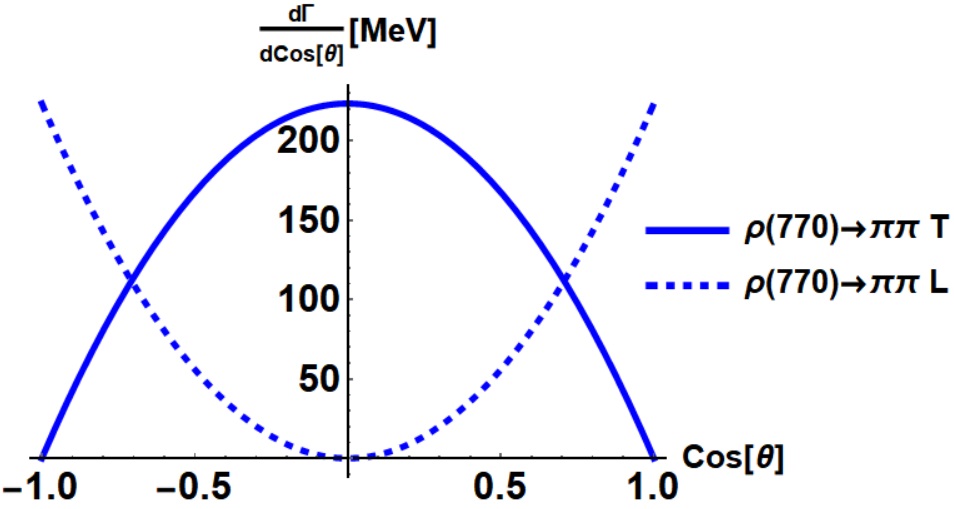}
\label{fig:rhodecay}
}

\subfigure [$\dv{\Gamma}{\cos\theta}$ of $K^{*}\to K\pi$ in the c.m. frame]{
\includegraphics[width=3.3in,height=1.7in]{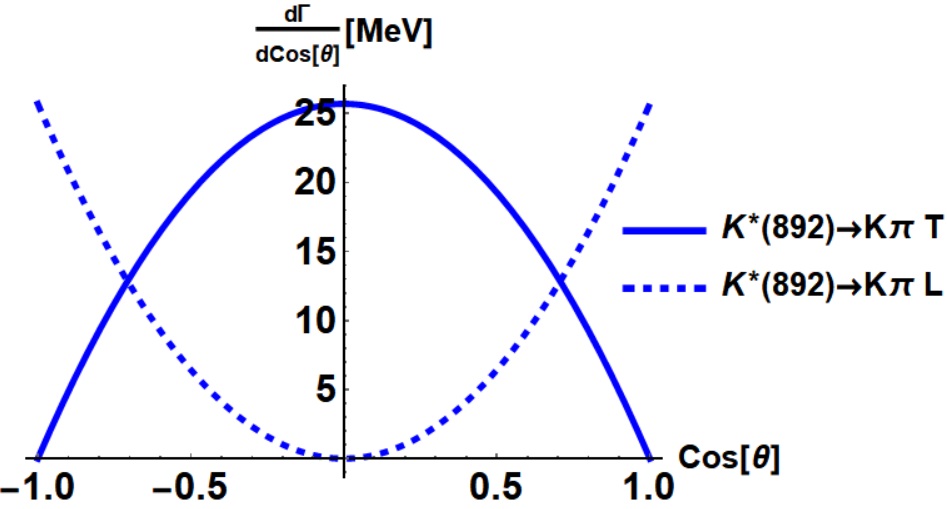}
\label{fig:kstardecay}
}
\caption{Angular distribution of decay rate of (a) $\rho\to\pi\pi$ and (b) $K^{*}\to K\pi$ in the c.m. frame for each polarization. T stands for transverse polarization and L stands for longitudinal polarization of the initial vector-meson.}
\label{fig:cos_theta_distribution1}
\end{figure}

\subsection{$K_{1}\to\rho K$ and $K_{1}\to K^{*}\pi$ decay}

The Lagrangian characterizing the coupling between the axial vector meson and a vector and pseudoscalar meson \cite{Sung:2021myr} is given as
\begin{align}
\mathcal{L}&=\sqrt{2}m_{K_{1}}\big(g_{K_{1}\rho K}\bar{K}\vec{\tau}\vdot\vec{\rho}_{\mu}-g_{K_{1}K^{*}\pi}\bar{K}^{*}_{\mu}\vec{\tau}\vdot\vec{\pi}\big)K_{1}^{\mu}.
\label{eqn:k1lagrangian}
\end{align}
The matrix representation of the $K_{1}$ field is given in Appendix\,\ref{appendix:effectivelagrangian}.

\begin{figure}
\hspace{2cm}
\includegraphics[width=3.4in,height=1.6in]{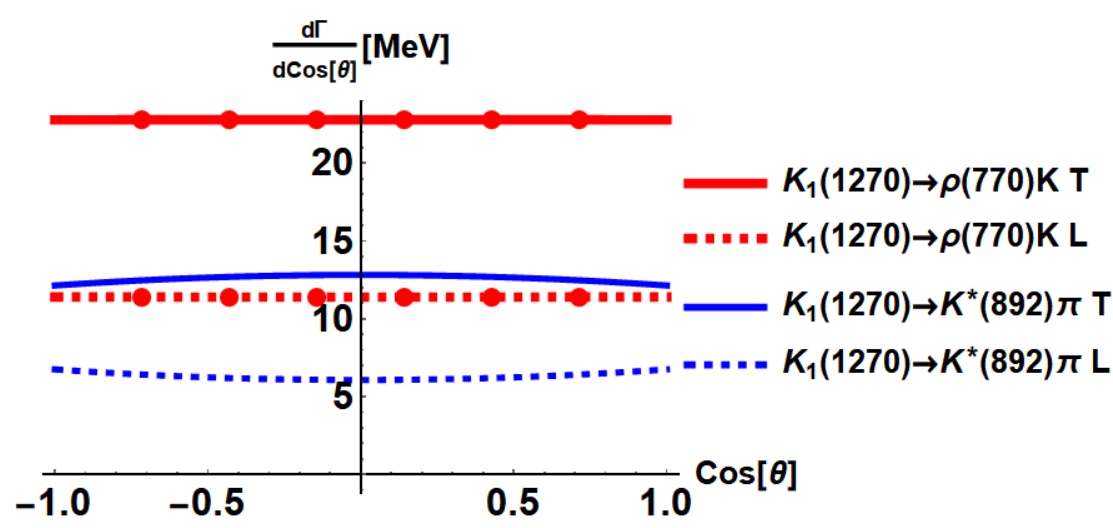}
\label{fig:rhokkstarpi}
\caption{Angular decay distribution of the $K_{1}\to\rho K$ and $K_{1}\to K^{*}\pi$ channels in the c.m. frame for an initially polarized $K_{1}$.}
\label{fig:cos_theta_distribution2}
\end{figure}

As before, we first compute the partial decay widths using the above interactions, giving
\begin{align}
\Gamma_{K_{1}\rho K}&=\frac{g_{K_{1}\rho K}^{2}}{8\pi}\abs{\bm{p_{1}}}\Bigg(3+\frac{\abs{\bm{p_{1}}}^{2}}{m_{\rho}^{2}}\Bigg)=34.2\,\text{MeV},\nonumber\\
&\\
\Gamma_{K_{1}K^{*}\pi}&=\frac{g_{K^{*}K\pi}^{2}}{8\pi}\abs{\bm{p_{1}}}\Bigg(3+\frac{\abs{\bm{p_{1}}}^{2}}{m_{K^{*}}^{2}}\Bigg)=18.9\,\text{MeV}.\nonumber
\end{align}
The partial decay widths of the $K_{1}$ decay channels are taken from the PDG \cite{Workman:2022ynf}.
Following the same procedure as in the previous subsection, the angular dependence of the decay distribution is obtained as
\begin{align}
&\frac{1}{\Gamma}\dv{\Gamma}{\Omega} \label{eq:K1_angular_decay}\\
&=\frac{3}{4\pi(3+\frac{\bm{p_{1}}^{2}}{m_{1}^{2}})}\Bigg(1+\frac{\bm{p_{1}^{2}}}{2m_{1}^{2}}\bigg(1-\rho_{00}+(3\rho_{00}-1)\cos^{2}\theta\nonumber\\
&-\sqrt{2}\text{Re}[\rho_{10}-\rho_{-10}]\sin2\theta\cos\phi\nonumber\\
&+\sqrt{2}\text{Im}[\rho_{10}+\rho_{-10}]\sin2\theta\sin\phi\nonumber\\
&-2\text{Re}[\rho_{1-1}]\sin^{2}\theta\cos2\phi+2\text{Im}[\rho_{1-1}]\sin^{2}\theta\sin2\phi\bigg)\Bigg).\nonumber
\end{align}
Integrating over $\phi$, we again get the polar angle $\theta$ distribution 
\begin{align}
&W(\theta) \label{eq:Wtheta2}\\
&=\frac{3}{2(3+\frac{\bm{p_{1}^{2}}}{m_{1}^{2}})}\Bigg(1+\frac{\bm{p_{1}^{2}}}{2m_{1}^{2}}\bigg(1-\rho_{00}+(3\rho_{00}-1)\cos^{2}\theta\bigg)\Bigg).\nonumber
\end{align}
The angular dependence of this distribution is shown in Fig.\,\ref{fig:cos_theta_distribution2}.  Unfortunately, unlike the case shown in Fig.\,\ref{fig:cos_theta_distribution1}, one can not isolate the different initial $K_{1}$ polarization by looking at different decay angles, 
which can be understood from the suppression factor $\bm{p_{1}^{2}}/(2m_{1}^{2}) = 6\times10^{-4}$ ($\rho K),=5.6\times10^{-2}$ ($K^{*} \pi$) appearing in the 
second term in the large bracket of Eq.\,(\ref{eq:Wtheta2}).

\begin{figure}
\centering
\subfigure [$\dv{\Gamma}{\cos\theta}$ of $K_{1}\to\rho K$ in the c.m. frame]{
\includegraphics[width=3.3in,height=1.6in]{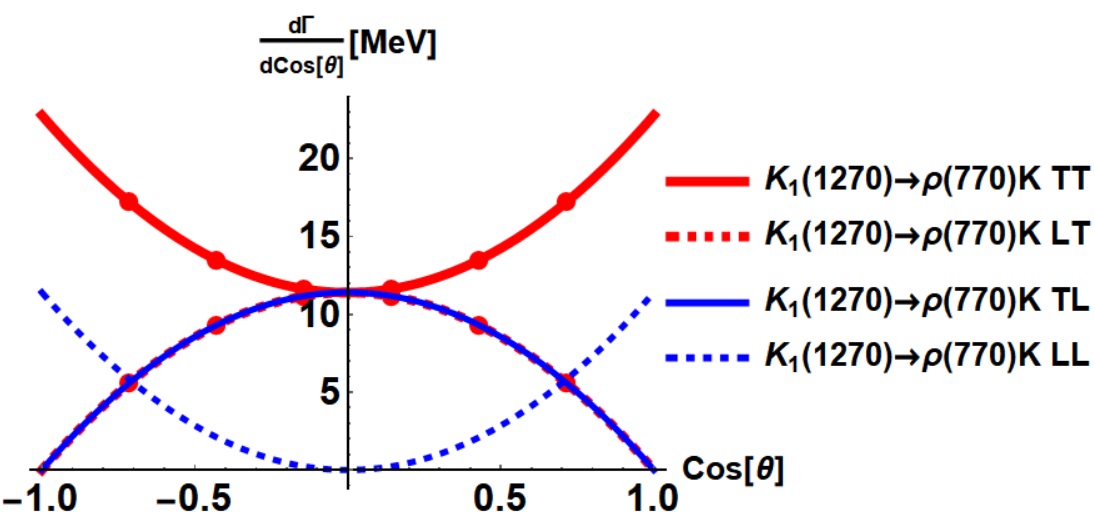}
\label{fig:k1rhokdecay}
}

\hspace{2cm}
\subfigure [$\dv{\Gamma}{\cos\theta}$ of $K_{1}\to K^{*}\pi$ in the c.m. frame]{
\includegraphics[width=3.3in,height=1.6in]{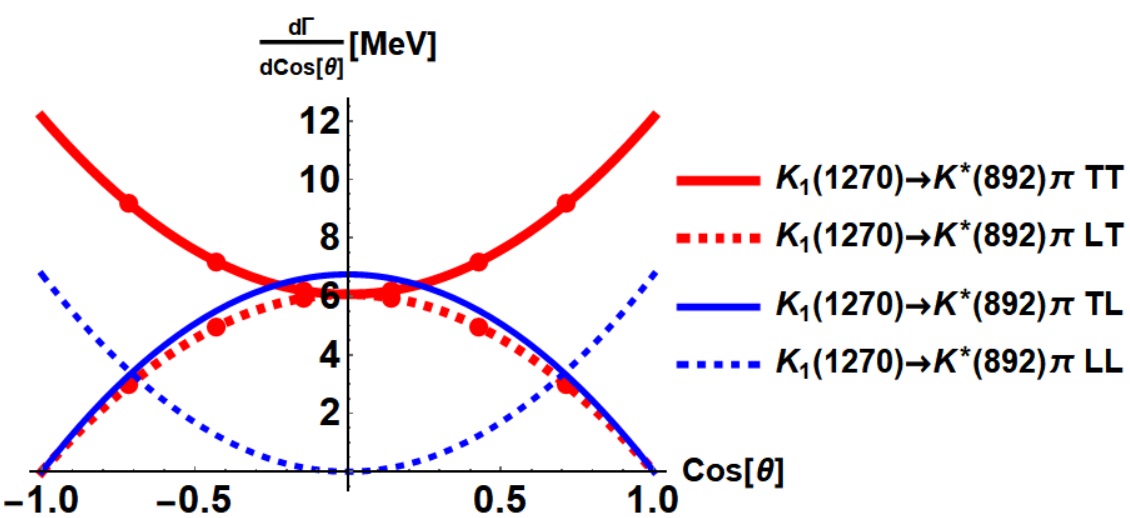}
\label{fig:k1kstarpidecay}
}
\caption{Angular distribution of decay rate of (a) $K_{1}\to\rho K$ and (b) $K_{1}\to K^{*}\pi$ in the c.m. frame for each polarization. The first and second T/L stands for the initial and final polarization of the spin-1 particle, respectively.}
\label{fig:cos_theta_distribution3}
\end{figure}

To overcome this, it is however possible to measure the polarization of the final vector-meson by again using the angular distribution as shown in Fig.\,\ref{fig:cos_theta_distribution1}.  Then, there are a total of four possible combinations of initial and final vector-meson polarizations. Each decay amplitude is listed below.
\begin{align}
\abs{\mathcal{M}_{TT}}^{2}&=2m_{K_{1}}^{2}g_{K_{1}VP}^{2}(1+\cos^{2}\theta),\nonumber\\
\abs{\mathcal{M}_{LT}}^{2}&=2m_{K_{1}}^{2}g_{K_{1}VP}^{2}\sin^{2}\theta,\nonumber\\
\abs{\mathcal{M}_{TL}}^{2}&=2m_{K_{1}}^{2}g_{K_{1}VP}^{2}\frac{E_{V}^{\prime2}}{m_{V}^{2}}\sin^{2}\theta,\\
\abs{\mathcal{M}_{LL}}^{2}&=2m_{K_{1}}^{2}g_{K_{1}VP}^{2}\frac{E_{V}^{\prime2}}{m_{V}^{2}}\cos^{2}\theta.\nonumber
\end{align}
The first and second subscripts of $\mathcal{M}$ ($T$ or $L$) here represent the polarization of an initial $K_{1}$ and final vector-meson, respectively. The corresponding results are shown in Fig.\,\ref{fig:cos_theta_distribution3}. As can be seen there, once we measure the transverse component of the final vector meson for both decay modes, one can isolate the transverse $K_1$ component by looking at the forward or backward direction. Conversely, when we measure the longitudinal component of the final vector meson, one can isolate the longitudinal $K_1$ component by again looking at the forward or backward direction. The improvement compared to the situation shown in 
Fig.\,\ref{fig:cos_theta_distribution2} is clear.

\subsection{Helicity basis and Wigner $D$-matrix}

So far, we have computed the general angular decay distribution by using the respective interaction Lagrangian for each decay channel. Here, we shall see that the same angular distribution is reproduced by taking advantage of the helicity formalism. As the basic ingredient, we 
need the Wigner $D$-matrix and the density matrix of an initial (axial)vector-meson. The convention for the Wigner $D$-matrix is adopted from that of Ref.\,\cite{Chung:1971ri} and \cite{devanathan2005angular}. The helicity basis for a massive particle is labeled by its momentum $\bm{p}$ and helicity 
$\lambda$ and is obtained by a boost along $z$-direction from the rest state followed by the rotation described by an Euler angle $(\phi,\theta,0)$. 
\begin{align}
&\mathcal{D}^{1}_{mm^{\prime}}(\phi,\theta,0)\nonumber\\
&=\begin{pmatrix} \frac{1+\cos\theta}{2}e^{-i\phi} & -\frac{1}{\sqrt{2}}\sin\theta e^{-i\phi} & \frac{1-\cos\theta}{2}e^{-i\phi} \\ \frac{1}{\sqrt{2}}\sin\theta & \cos\theta & -\frac{1}{\sqrt{2}}\sin\theta \\ \frac{1-\cos\theta}{2}e^{i\phi} & \frac{1}{\sqrt{2}}\sin\theta e^{i\phi} & \frac{1+\cos\theta}{2} e^{i\phi}\end{pmatrix},
\label{eq:spin1rotation_matrix}
\end{align}
By applying a Wigner $D$-matrix, we rotate the density matrix so that the quantization axis rotates from the $z$-axis to align with the direction of momentum of an outgoing particle, specified by the angles $\phi$ and $\theta$.
\begin{align}
\Bigg(\frac{1}{\Gamma}\dv{\Gamma}{\Omega}\Bigg)=\frac{3}{4\pi}\sum_{\lambda,m,m^{\prime}=\pm1,0}D^{1\dagger}_{\lambda m}\rho_{m m^{\prime}}D^{1}_{m^{\prime}\lambda}\abs{H(\lambda_{1},\lambda_{2})}^{2}, 
\label{eq:Wigner_D_Matrix}
\end{align}
Here, $\lambda=\lambda_{1}-\lambda_{2}$, where $\lambda_{1}$ and $\lambda_{2}$ are the helicities of the daughter particles of mass $m_{1}$ and $m_{2}$, respectively.  $H(\lambda_{1},\lambda_{2})$ here stands for the interaction Hamiltonian for each helicity component of corresponding decay, which we can calculate from the interaction Lagrangians given before. More details regarding this calculation are explained in  Appendix\,\ref{appendix:appendixb}.

For example, for the  $V \rightarrow PP$ decay, since the final particles are spinless, one should use Eq.\,(\ref{eq:Wigner_D_Matrix}) with only the $\lambda=0$ component by definition. This exactly reproduces the result of  Eq.\,(\ref{eq:vpp_angle_dist}). For $A \rightarrow VP$, one on the other hand needs to sum over all the helicity components $\lambda = \pm1$, 0.

\section{Summary and Conclusions}\label{sec:sumandconc}

In this work, we have shown that one can isolate the initial longitudinal and transverse modes of the $K^*$ and $K_1$ from observing the decay angles and polarizations of their decay particles. In particular, for $K^*$, this is possible by measuring the decay angle distributions of the outgoing pseudoscalar mesons. For $K_1$, one furthermore needs to determine the polarization of the outgoing vector meson to disentangle the longitudinal and transverse modes.

Such a measurement should be feasible in a future J-PARC experiment. This will help to reduce the uncertainty of the mass shift measurement of these two particles in nuclear matter. Once this is realized, the chiral partner nature of $K^*$ and $K_1$ may be experimentally confirmed, which will bring us one step closer to understanding the role of chiral symmetry breaking and restoration to the generation of hadron masses.

\section*{Acknowledgments}

This work was supported by the Samsung Science and Technology Foundation under Project 
No. SSTF-BA1901-04, 
and by the Korea National Research Foundation under grant No. 2023R1A2C3003023 and No. 2023K2A9A1A0609492411, 
JSPS KAKENHI Grants No. JP19KK0077, No. JP20K03940, No. JP21H00128, 
No. JP21H00128, and No. JP21H01102, 
for the Promotion of Science (JSPS). The work was also supported by the REIMEI project of JAEA under the title ``Studying the origin of hadron masses through the behavior of vector mesons in nuclear matter from theory and experiment".

\clearpage
\appendix

\renewcommand{\thesection}{\Alph{section}}
\renewcommand{\thesubsection}{\Alph{section}-\arabic{subsection}}

\begin{widetext}

%\section{Effective interaction Lagrangian and polarization tensor of a vector meson}\label{appendix:appendixa}
\section{Effective interaction Lagrangian}\label{appendix:effectivelagrangian}

%\subsection{Effective interaction Lagrangian}\label{appendix:effectivelagrangian}

$K$, $K^{*}$ and $K_{1}$ isodoublet matrices are defined as
\begin{align}
K^{*}&=\begin{pmatrix}K^{+*}\\K^{0*}\end{pmatrix},\;\bar{K}=\begin{pmatrix}K^{-}&\bar{K}^{0}\end{pmatrix},\,K_{1}=\begin{pmatrix}K_{1}^{+}\\K_{1}^{0}\end{pmatrix},\,\bar{K}^{*}=\begin{pmatrix}K^{-*}&\bar{K}^{0*}\end{pmatrix}.
\end{align}
Direct matrix multiplication yields the interaction Lagrangian as
\begin{align}
\mathcal{L}_{\rho\pi\pi}&=g_{\rho\pi\pi}\bigg((\pi^{+}\partial_{\mu}\pi^{-}-\partial_{\mu}\pi^{+}\pi^{-})\rho^{0\mu}+(\pi^{-}\partial_{\mu}\pi^{0}-\partial_{\mu}\pi^{-}\pi^{0})\rho^{+\mu}+(\pi^{+}\partial_{\mu}\pi^{0}-\partial_{\mu}\pi^{+}\pi^{0})\rho^{-\mu}\bigg),\\
\mathcal{L}_{K^{*}K\pi}&=ig_{K^{*}K\pi}\bigg[\sqrt{2}\bigg(K^{-}\partial_{\mu}\pi^{0}-\partial_{\mu}K^{-}\pi^{0}+\bar{K}^{0}\partial_{\mu}\sqrt{2}\pi^{-}-\partial_{\mu}\bar{K}^{0}\sqrt{2}\pi^{-}\bigg)K^{+*\mu}\\
&+\sqrt{2}\bigg(K^{-}\partial_{\mu}\sqrt{2}\pi^{+}-\partial_{\mu}K^{-}\sqrt{2}\pi^{+}-\bar{K}^{0}\partial_{\mu}\pi^{0}+\partial_{\mu}\bar{K}^{0}\pi^{0}\bigg)K^{0*\mu}\bigg],\nonumber\\
\mathcal{L}_{K_{1}VP}&=im_{K_{1}}g_{K_{1}K^{*}\pi}\sqrt{2}\bigg[\bigg(K^{-*}_{\mu}\pi^{0}+\bar{K}^{0*}_{\mu}\sqrt{2}\pi^{-}\bigg)K_{1}^{+\mu}+\bigg(K^{-*}_{\mu}\sqrt{2}\pi^{+}-\bar{K}^{0*}_{\mu}\pi^{0}\bigg)K_{1}^{0\mu}\bigg]\\
&-im_{K_{1}}g_{K_{1}\rho K}\sqrt{2}\bigg[\bigg(K^{-}\rho^{0}_{\mu}+\bar{K}^{0}\sqrt{2}\rho^{+}_{\mu}\bigg)K_{1}^{+\mu}+\bigg(K^{-}\sqrt{2}\rho^{+}_{\mu}-\bar{K}^{0}\rho^{0}_{\mu}\bigg)K_{1}^{0\mu}\bigg].\nonumber
\end{align}

%\subsection{Disentangling the polarizations of a vector-meson using polarization tensor}\label{appendix:polarizationtensor}
\section{Disentangling the polarizations of a vector-meson using the polarization tensor and vector}\label{appendix:polarizationtensor}
We will in this appendix discuss two methods to disentangle the contributions of different polarization components of vector 
mesons to their decay amplitudes. The first one makes use of the polarization tensor and can only be used for purely transversely or 
longitudinally polarized vector/axial-vector mesons. The second more general method uses the polarization vector and can be applied to an arbitrary 
spin configuration.
For the first method, we first need to define the polarization tensors as 
\begin{align}
P_{T}^{\mu\nu}&=\begin{pmatrix}0&0\\0&\delta^{ij}-\frac{q^{i}q^{j}}{\bm{q}^{2}}\end{pmatrix},\,P_{L}^{\mu\nu}=\begin{pmatrix}\frac{\bm{q}^{2}}{m_{v}^{2}}&\frac{E_{v}q^{i}}{m_{v}^{2}}\\\frac{E_{v}q^{i}}{m_{v}^{2}}&\frac{E_{v}^{2}q^{i}q^{j}}{m_{v}^{2}\bm{q}^{2}}\end{pmatrix}.
\end{align}
Contracting these polarization tensors with the decay amplitude, we can disentangle its transverse and longitudinal parts. $q_{0}$ and $q^{i}$ here stand for the energy and momentum of the considered vector/axial-vector meson. 
In what follows, we will display the $\rho\to\pi\pi$ decay as an example of $V\to PP$ and the $K_{1}\to\rho K$ decay as an example of $A\to VP$. The same method can also be applied to $K^{*}\to K\pi$ and $K_{1}\to K^{*}\pi$, respectively. 
The decay amplitudes for the two cases are obtained as
\begin{align}
\mathcal{M}_{\mu\nu}&=g_{\rho\pi\pi}^{2}(p_{1}-p_{2})_{\mu}(p_{1}-p_{2})_{\nu},\\
\mathcal{M}_{\mu\nu}&=2m_{K_{1}\rho K}^{2}g_{K_{1}\rho K}^{2}\varepsilon_{\mu}(\lambda_{\rho})\varepsilon_{\nu}^{*}(\lambda_{\rho}).\nonumber
\end{align}
Contracting these with the above polarization tensors, taking the final spin sum (if applicable), we get
\begin{align}\label{eqn:decayamplitude}
\rho\to\pi+\pi&\begin{cases}\abs{\mathcal{M}}_{T}^{2}&=2g_{\rho\pi\pi}^{2}\bm{p_{1}}^{2}\sin^{2}\theta,\\\abs{\mathcal{M}}_{L}^{2}&=4g_{\rho\pi\pi}^{2}\bm{p_{1}}^{2}\cos^{2}\theta,\end{cases}
\end{align}
and
\begin{align}\label{eqn:decayamplitude2}
K_{1}\to\rho+K&\begin{cases}\abs{\mathcal{M}}_{T}^{2}=m_{K_{1}}^{2}g_{K_{1}\rho K}^{2}\Bigg(2+\frac{\bm{p_{1}}^{2}}{m_{\rho}^{2}}\sin^{2}\theta\Bigg),\\\abs{\mathcal{M}}_{L}^{2}=2m_{K_{1}}^{2}g_{K_{1}\rho K}^{2}\Bigg(1+\frac{\bm{p_{1}}^{2}}{m_{\rho}^{2}}\cos^{2}\theta\Bigg).\end{cases}
\end{align}
The different factors 2 and 1 appearing in the fist terms within the large brackets in Eq.\,(\ref{eqn:decayamplitude2}) are due to the different degeneracy factors of the two transverse and one longitudinal modes for a massive spin-1 particle.

Let us next move on to the second method, in which we can further study the contributions of the different helicity states and their mixing. The polarization vectors of the initial particle in its own rest frame for each helicity state are given as
\begin{align}
\varepsilon^{\mu}(0,\pm1)&=\begin{pmatrix}0\\\mp\frac{1}{\sqrt{2}}\\-\frac{i}{\sqrt{2}}\\0\end{pmatrix},\,\,\,\,\varepsilon^{\mu}(0,0)=\begin{pmatrix}0\\0\\0\\1\end{pmatrix},
\end{align}
where 
$\varepsilon^{\mu}(\bm{p},\lambda)$ is the general polarization vector (that will be more explicitly discussed further below) 
with $\bm{p}$ being the particle momentum and $\lambda$ its helicity.
Taking the absolute square of the invariant amplitude
\begin{align}
\mathcal{M}_{VPP}&=g_{VPP}(p_{1}-p_{2})_{\mu}\sum_{\lambda_{V}=\pm1,0}a_{\lambda_{V}}\varepsilon^{\mu}(\lambda_{V}),
\end{align}
yields the general angular distribution which can be expressed as
\begin{align}
|\mathcal{M}|^{2}&=2g_{VPP}^{2} |\bm{p}_1|^2 
\Bigg(1-\rho_{00}+(3\rho_{00}-1)\cos^{2}\theta-2\text{Re}[\rho_{1-1}]\sin^{2}\theta\cos2\phi+2\text{Im}[\rho_{1-1}]\sin^{2}\theta\sin2\phi\nonumber\\
&-\sqrt{2}\text{Re}[\rho_{10}-\rho_{-10}]\sin2\theta\cos\phi+\sqrt{2}\text{Im}[\rho_{10}+\rho_{-10}]\sin2\theta\sin\phi\Bigg), 
\end{align}
where $\rho_{\lambda\lambda^{\prime}}$ is defined in Eq.\,(\ref{eq:spin_density_def}).

For the $A\to VP$ decay, we also need the polarization vector of the produced vector-meson in the rest frame of the initial axial vector-meson, which is obtained by an inverse Lorentz boost along $z$-axis followed by an Euler rotation $R(\phi,\theta,0)$. $R(\alpha,\beta,\gamma)$ here rotates the object about the  $z$-axis by an angle of $\gamma$, followed by a rotation around the  $y$-axis by an angle of $\beta$, and finally followed by an angle of $\alpha$ around the  $z$-axis. The polarization vectors of the produced vector-meson in the c.m. frame are then obtained as
\begin{align}
&\varepsilon^{\mu}(\vec{p},\pm1)=\begin{pmatrix}0\\\mp\frac{1}{\sqrt{2}}\cos\theta\cos\phi+\frac{i}{\sqrt{2}}\sin\phi\\\mp\frac{1}{\sqrt{2}}\cos\theta\sin\phi-\frac{i}{\sqrt{2}}\cos\phi\\\pm\frac{1}{\sqrt{2}}\sin\theta\end{pmatrix},\,\varepsilon^{\mu}(\vec{p},0)=\begin{pmatrix}\frac{\abs{\bm{p_{1}}}}{m}\\\frac{E^{\prime}}{m}\sin\theta\cos\phi\\\frac{E^{\prime}}{m}\sin\theta\sin\phi\\\frac{E^{\prime}}{m}\cos\theta\end{pmatrix}.
\end{align}
$E_{1}^{\prime}$ here is the energy of the produced vector-meson in the c.m. frame. The general angular distribution is calculated as
\begin{align}
\mathcal{M}_{K_{1}VP}(\lambda_{V})&=\sqrt{2}m_{K_{1}}g_{K_{1}VP}\sum_{\lambda_{K_{1}}=\pm1,0}a_{\lambda_{K_{1}}}\varepsilon^{\mu}(\lambda_{K_{1}})\varepsilon^{*}_{\mu}(\lambda_{V}),\\
\sum_{\lambda_{V}=\pm1,0}\abs{\mathcal{M}_{K_{1}VP}(\lambda_{V})}^{2}&=2m_{K_{1}}^{2}g_{K_{1}VP}^{2}\Bigg(1+\frac{\bm{p_{1}^{2}}}{2m_{1}^{2}}\bigg(1-\rho_{00}+(3\rho_{00}-1)\cos^{2}\theta-2\text{Re}[\rho_{1-1}]\sin^{2}\theta\cos2\phi\\
&+2\text{Im}[\rho_{1-1}]\sin^{2}\theta\sin2\phi-\sqrt{2}\text{Re}[\rho_{10}-\rho_{-10}]\sin2\theta\cos\phi+\sqrt{2}\text{Im}[\rho_{10}+\rho_{-10}]\sin2\theta\sin\phi\bigg)\Bigg)\nonumber.
\end{align}

\section{More details about the helicity formalism}\label{appendix:appendixb}
The two-body decay process is considered starting from a definite angular momentum state of $\ket{JM}$ in the mother particle rest frame, 
decaying into two particle helicity state $\ket{p\phi\theta\lambda}$, where $p=p_1-p_2$ and  $\lambda=\lambda_1-\lambda_2$ are the relative momenta and helicity difference, respectively, between the two decaying particles denoted with subscripts 1 and 2. The notation and derivation of this section are adapted from 
Ref.\,\cite{Chung:1971ri}.

\subsection{One particle state}\label{appendix:oneparticle}
First, we study the single-particle canonical and helicity states. The canonical state is defined as a state labeled by its momentum, total angular momentum and its $z$-component.  
A general canonical state with arbitrary momentum pointing in the $\phi$, $\theta$ direction is then constructed by first inversely rotating the particle such that it aligns with the z-axis, followed by a Lorentz boost in the z-direction, and finally a rotation back into the momentum direction of the particle with polar angles $(\phi,\theta)$,
\begin{align}
\ket{\vec{p}jm}&=U\big(R(\phi,\theta,0)L_{z}(p)R^{-1}(\phi,\theta,0)\big)\ket{0jm},
\end{align}
where $L_{z}(p)$ is a Lorentz boost along the $z$-axis. 
When the particle is at rest, the canonical state  transforms under rotation as
\begin{align}
U(R)\ket{0jm}&=\sum_{m^{\prime}}D^{j}\tensor{(R)}{_{m^{\prime}}_{m}}\ket{0jm^{\prime}}, 
\label{eq:rotation_property}
\end{align}
where $D^{j}(R)_{m^{\prime}m}$ is a linear representation of a rotation operator $U(R)$ \cite{Chung:1971ri}.

The helicity state is labeled by the momentum, total angular momentum and helicity. It is similarly constructed by firstly Lorentz boosting the 
rest state $\ket{0j\lambda}$ (which is here defined such that $\lambda$ is the eigenstate of the z-component of the angular momentum, it is thus the same as $\ket{0jm}$ with $m = \lambda$). along the z-axis followed by a rotation such that the momentum points into the 
direction specified by the polar angles $\phi$ and $\theta$. We thus have 
\begin{align}
\ket{\vec{p}j\lambda}&=U\big(R(\phi,\theta,0)L_{z}(p) \big)\ket{0j\lambda}.%\\
%U(R)\ket{\vec{p}j\lambda}&=U(R)U\big(R(\phi,\theta,0)L_{3}(p)\big)\ket{0j\lambda}=\ket{R\vec{p}j\lambda}
\end{align}
The relation between canonical and helicity states is given as 
\begin{align}
\ket{\vec{p}j\lambda}&=U\big(R(\phi,\theta,0)L_{z}(p)R^{-1}(\phi,\theta,0)\big)U\big(R(\phi,\theta,0)\big)\ket{0j\lambda}=\sum_{m}D^{j}(\phi,\theta,0)_{m\lambda}\ket{\vec{p}jm}
\end{align}
We here choose our normalization to be Lorentz invariant, such that

\begin{align}
\braket{\vec{p}^{\prime}j^{\prime}\lambda^{\prime}}{\vec{p}j\lambda}&=(2\pi)^{3}2E_{\vec{p}}\delta^{3}(\vec{p}-\vec{p}^{\prime})\delta_{jj^{\prime}}\delta_{\lambda\lambda^{\prime}},\quad\braket{\vec{p}^{\prime}j^{\prime}m^{\prime}}{\vec{p}jm}=(2\pi)^{3}2E_{\vec{p}}\delta^{3}(\vec{p}-\vec{p}^{\prime})\delta_{jj^{\prime}}\delta_{mm^{\prime}}.
\end{align}

\subsection{Two particle state}\label{appendix:twoparticle}
By definition, the two particle helicity state is a tensor product of two one particle states in the c.m. frame,  
\begin{align}
\ket{\phi\theta\lambda_{1}\lambda_{2}}&=U\big(R(\phi,\theta,0)\big)\big[U\big(L_{z}(p)\big)\ket{0j_{1}\lambda_{1}}\otimes U\big(L_{z}(-p)\big)\ket{0j_{2}-\lambda_{2}}\big].
\end{align}
We next derive the relation between the two particle helicity state and a state of definite angular momentum $\ket{JM\lambda_{1}\lambda_{2}}$.  Here, $J$ and $M$ denote the total angular momentum and its projection onto the $z$-axis of the initial particle, respectively. We assume that the above general two particle helicity state, with the momentum of one particle specified by the angles $(\phi,\theta)$ in the c.m. frame, is related to the total angular momentum state by a coefficient $C_{JM}(\phi,\theta,\lambda_{1},\lambda_{2})$ as
\begin{align}
\ket{\phi\theta\lambda_{1}\lambda_{2}}&=\sum_{JM}C_{JM}\big(\phi,\theta,\lambda_{1},\lambda_{2}\big)\ket{JM\lambda_{1}\lambda_{2}}.
\label{eq:derive_C_1}
\end{align}
Let us here derive an explicit expression for $C_{JM}(\phi,\theta,\lambda_{1},\lambda_{2})$. The standard helicity state is defined for the state where $\phi=\theta=0$, 
\begin{align}
\ket{00\lambda_{1}\lambda_{2}}&=\sum_{JM}C_{JM}(0,0,\lambda_{1},\lambda_{2})\ket{JM\lambda_{1}\lambda_{2}}=\sum_{J}C_{J\lambda}(0,0,\lambda_{1},\lambda_{2})\ket{J\lambda\lambda_{1}\lambda_{2}},
\end{align}
In the standard state, particle 2 is heading towards the negative $z$-direction, thus its projection on the $z$-axis is $-\lambda_{2}$. Therefore, total angular momentum projection $\lambda=\lambda_{1}-\lambda_{2}$. 
By a definition of two particle helicity state, it can also be viewed as a state which is rotated from the standard state. 
Hence, 
\begin{align}
\ket{\phi\theta\lambda_{1}\lambda_{2}}&=U(R)\ket{00\lambda_{1}\lambda_{2}}=\sum_{J}C_{J\lambda}(0,0,\lambda_{1},\lambda_{2})U(R)\ket{J\lambda\lambda_{1}\lambda_{2}}\nonumber\\
&=\sum_{JM}C_{J\lambda}(0,0,\lambda_{1},\lambda_{2})D^{J}(R)_{M\lambda}\ket{JM\lambda_{1}\lambda_{2}}, 
\label{eq:derive_C_2}
\end{align}
where we have in the last line made use of the fact that a state of definite angular momentum behaves the same way as given in 
Eq.\,(\ref{eq:rotation_property}). 
Making use of proper orthogonality relations of the states $\ket{\phi\theta\lambda_{1}\lambda_{2}}$ and $\ket{JM\lambda_{1}\lambda_{2}}$ 
and properties of the rotation matrix $D^{J}(R)_{M\lambda}$ (see for example Ref.\,\cite{Chung:1971ri} for more details), we obtain 
$C_{JM}(0,0,\lambda_{1},\lambda_{2})$ as
\begin{align}
C_{JM}(0,0,\lambda_{1},\lambda_{2}) = \sqrt{\frac{2J+1}{4\pi}}, 
\end{align}
and, comparing Eq.\,(\ref{eq:derive_C_1}) with the last line of Eq.\,(\ref{eq:derive_C_2}), 
we finally have 
\begin{align}
C_{JM}(\phi,\theta,\lambda_{1},\lambda_{2}) = \sqrt{\frac{2J+1}{4\pi}} D^{J}(R)_{M\lambda}, 
\label{eq:derive_C_3}
\end{align}
where again $\lambda = \lambda_{1}-\lambda_{2}$.

\subsection{Two body decay amplitude}\label{appendix:decayamplitude}
The two body decay amplitude is a transition amplitude from a definite angular momentum state $\ket{JM}$ of the initial particle to a two particle helicity state $\ket{\phi\theta\lambda_{1}\lambda_{2}}$ of the daughter particles in the c.m. frame. The transition amplitude from $\ket{JM}$ to $\ket{\phi\theta\lambda_{1}\lambda_{2}}$ is given as below,
\begin{align}
f_{\lambda M}&=\mel{\phi\theta\lambda_{1}\lambda_{2}}{H_{\text{int}}}{JM}=\sum_{J'M'\lambda'_{1}\lambda'_{2}}\braket{\phi\theta\lambda_{1}\lambda_{2}}{J'M'\lambda_{1}'\lambda_{2}'}\mel{J'M'\lambda_{1}'\lambda_{2}'}{H_{\text{int}}}{JM}\nonumber\\
&=\sqrt{\frac{2J+1}{4\pi}}D^{J}(R)^{*}_{M\lambda}\mel{JM\lambda_{1}\lambda_{2}}{H_{\text{int}}}{JM}, 
\label{eq:transition_amplitude}
\end{align}
where Eq.\,(\ref{eq:derive_C_3}) and angular momentum conservation was used in the second line. 
$H_{\text{int}}$ here stands for the interaction Hamiltonian describing the decay. 
Making use of the fact that this is a scalar quantity, the matrix element $\mel{JM\lambda_{1}\lambda_{2}}{H_{\text{int}}}{JM}$ 
cannot depend on $M$, but only on the rotational invariants $J$, $\lambda_{1}$ and $\lambda_{2}$. 
We will hence denote it as $\mel{JM\lambda_{1}\lambda_{2}}{H_{\text{int}}}{JM} \equiv H^{J}_{\text{int}}(\lambda_{1},\lambda_{2})$ in what follows. 

If we specify the initial state $\ket{I}$ as superposition of the different $M$ quantum numbers, specifically 
$\ket{I} = \sum_{M} a_{M}\ket{JM}$ and in analogy to Eq.\,(\ref{eq:spin_density_def}) define the 
spin density matrix as $\rho_{MM^{\prime}} = a_{M} a^{*}_{M^{\prime}}$, 
the normalized angular distribution of this decay can be given as 
$I(\phi,\theta) = |\sum_{M} a_{M} f_{\lambda M}|^2/\Gamma$, where $\Gamma$ is the decay width 
of the initial particle. We hence obtain 
\begin{align}
I(\phi,\theta)&=\frac{1}{\Gamma}f_{\lambda M}\rho_{MM^{\prime}}f^{*}_{\lambda M^{\prime}}=\frac{1}{\Gamma}\mel{\phi\theta\lambda_{1}\lambda_{2}}{H_{\text{int}}}{JM}\rho_{MM^{\prime}}\mel{JM^{\prime}}{H_{\text{int}}}{\phi\theta\lambda_{1}\lambda_{2}}\nonumber\\
&=\frac{1}{\Gamma}\sum_{\lambda_{1}\lambda_{2}}\sum_{MM^{\prime}}\frac{2J+1}{4\pi}D^{J\dagger}\tensor{(\phi,\theta,0)}{_\lambda_M}\rho_{MM^{\prime}}D^{J}\tensor{(\phi,\theta,0)}{_{M^{\prime}}_\lambda}\abs{H^{J}_{\text{int}}(\lambda_{1},\lambda_{2})}^{2}.
\end{align}
For the $\rho\to\pi\pi$ decay, only the matrix element $H^{1}_{\text{int}}(0,0)$ is needed, and the angular decay distribution 
is therefore automatically fixed only from the rotation matrix $D^{1}(\phi,\theta,0)$, given in Eq.\,(\ref{eq:spin1rotation_matrix}). 
As a result, we obtain
\begin{align}
I(\phi,\theta)&=\frac{3}{8\pi}\Big(1-\rho_{00}+(3\rho_{00}-1)\cos^{2}\theta-\sqrt{2}\text{Re}[\rho_{10}-\rho_{-10}]\sin2\theta\cos\phi+\sqrt{2}\text{Im}[\rho_{10}+\rho_{-10}]\sin2\theta\sin\phi\\
&-2\text{Re}[\rho_{1-1}]\sin^{2}\theta\cos2\phi+2\text{Im}[\rho_{1-1}]\sin^{2}\theta\sin2\phi\Big), \nonumber
\end{align}
which agrees with Eq.\,(\ref{eq:vpp_angle_dist}). 

On the other hand, for the $K_{1}\to\rho K$ decay, the three matrix elements $H^{1}_{\text{int}}(1,0)$, $H^{1}_{\text{int}}(0,0)$ 
and $H^{1}_{\text{int}}(-1,0)$ need to be considered. 
Confining us here to strong and thus parity conserving decay, we can make use of the symmetry 
property of $H^{1}_{\text{int}}(1,0) = H^{1}_{\text{int}}(-1,0)$ and are hence left with two independent terms, 
which have to be determined from a specific interaction Hamiltonian. 
In this work, it can be easily obtained from the interaction Lagrangian given in Eq.\,(\ref{eqn:k1lagrangian}) and $H_{\text{int}}=-L_{\text{int}}$.  
Next, we compute the transition amplitude of Eq.\,(\ref{eq:transition_amplitude}) using the polarization vectors given in Appendix B. 
To determine the relative strength of the two terms, we only need two independent transition amplitudes, with 
with an outgoing vector particle carrying a different helicity $\lambda_{1}$. 
Specifically, we have
\begin{align}
f_{1 1} \propto \epsilon^{\ast}_{\rho}(1) \cdot \epsilon_{K_1}(1) = -\frac{1 + \cos \theta}{2} e^{i\phi}, 
\label{eq:transition_amplitude_11}
\end{align}
and 
\begin{align}
f_{0 1} \propto \epsilon^{\ast}_{\rho}(0) \cdot \epsilon_{K_1}(1) = -\frac{E'}{\sqrt{2}m} \sin \theta e^{i\phi}.  
\label{eq:transition_amplitude_01}
\end{align}
Comparing this with Eq.\,(\ref{eq:transition_amplitude}), we note that 
\begin{align}
H^{1}_{\text{int}}(1,0) = H^{1}_{\text{int}}(-1,0) \propto g_{K_{1}\rho K},
\end{align}
and
\begin{align}
H^{1}_{\text{int}}(0,0) \propto g_{K_{1}\rho K}\frac{E'}{m}.
\end{align}
This is sufficient to derive the angular distribution of the $K_{1}\to\rho K$ decay as
\begin{align}
I(\phi,\theta)&=\frac{1}{\Gamma}\frac{3}{4\pi}\Bigg(\abs{H(1,0)}^{2}+\frac{1}{2}\Big(\abs{H(0,0)}^{2}-\abs{H(1,0)}^{2}\Big)\nonumber\\
&\Big(1-\rho_{00}+(3\rho_{00}-1)\cos^{2}\theta-\sqrt{2}\text{Re}[\rho_{10}-\rho_{-10}]\sin2\theta\cos\phi+\sqrt{2}\text{Im}[\rho_{10}+\rho_{-10}]\sin2\theta\sin\phi\nonumber\\
&-2\text{Re}[\rho_{1-1}]\sin^{2}\theta\cos2\phi+2\text{Im}[\rho_{1-1}]\sin^{2}\theta\sin2\phi\Big)\Bigg)\\
&=\frac{3}{4\pi\big(3+\frac{\bm{p_{1}}^{2}}{m_{1}^{2}}\big)}\bigg(1+\frac{\bm{p_{1}}^{2}}{2m_{1}^2}\Big(1-\rho_{00}+(3\rho_{00}-1)\cos^{2}\theta-\sqrt{2}\text{Re}[\rho_{10}-\rho_{-10}]\sin2\theta\cos\phi\nonumber\\
&+\sqrt{2}\text{Im}[\rho_{10}+\rho_{-10}]\sin2\theta\sin\phi-2\text{Re}[\rho_{1-1}]\sin^{2}\theta\cos2\phi+2\text{Im}[\rho_{1-1}]\sin^{2}\theta\sin2\phi\Big)\bigg), \nonumber
\end{align}
which agrees with Eq.\,(\ref{eq:K1_angular_decay}).

\end{widetext}

\end{document}